\documentclass[preprint,12pt]{elsarticle}

\usepackage[table]{xcolor}
\usepackage[labelsep=quad]{subfig}

\usepackage{amsmath,epsfig}
\usepackage{colortbl}
\usepackage{graphicx}
\usepackage{amscd}
\usepackage{amssymb}
\usepackage{booktabs}
\usepackage{multirow}
\usepackage{stfloats}
\usepackage{bm,xstring}
\usepackage{float}

\journal{Signal Processing}

\begin{document}

\begin{frontmatter}

\title{A Topological Loss Function for Image Denoising on a new BVI-lowlight Dataset}

\affiliation[inst1]{organization={Visual Information Laboratory, University of Bristol},
           addressline={1 Cathedral Square}, 
            city={Bristol},
            postcode={BS8 1UB}, 
            country={UK}}
\author[inst1]{Alexandra Malyugina}

\author[inst1]{Nantheera Anantrasirichai}

\author[inst1]{David Bull}

\begin{abstract}

Although image denoising algorithms have attracted significant research attention, surprisingly few have been proposed for, or evaluated on,  noise from imagery acquired under real low-light conditions. Moreover, noise characteristics are often assumed to be spatially invariant, leading to edges and textures being distorted after denoising. Here, we introduce a novel topological loss function which is based on persistent homology. The method performs in the space of image patches, where topological invariants are calculated and represented in persistent diagrams. The loss function is a combination of $\ell_1$ or $\ell_2$ losses with the new persistence-based topological loss. We compare its performance across popular denoising architectures and loss functions, training the networks on our new comprehensive dataset of natural images captured in low-light conditions -- BVI-LOWLIGHT. Analysis reveals that this approach outperforms existing methods, adapting well to complex structures and suppressing common artifacts.


\end{abstract}
%

\begin{highlights}
\item Image denoising suffers from lack of comprehensive datasets -  specifically, those captured in low light conditions
\item Our BVI-lowlight dataset \cite{bvilowlight} can be used for training and benchmarking supervised denoising algorithms on images with a wide range of realistic ISO noise levels
\item Incorporating topological properties of clean/noisy images into the loss function improves denoising performance by both increasing objective metrics and enhancing subjective results
\end{highlights}

\begin{keyword}

image denoising \sep  TDA \sep image dataset \sep loss function \sep  persistent homology  

\end{keyword}

\end{frontmatter}


\section{Introduction}
\label{sec:intro}

The management of noise is essential in any image processing pipeline whether it be to improve computer vision tasks like classification in scientific or medical images, object detection and tracking in self-driving in poor light conditions, or to deliver more immersive visual experiences in consumer videos. Denoising is generally considered an ill-posed inverse problem. It can be defined as the process of mapping from a noisy signal to a noise-free version where the noise is removed while the underlying signal is preserved. However, in practical situations, it is impossible to perfectly restore the underlying noise-free signal. Hence, denoising is always a trade-off between signal distortion and noise removal.
There do exist self-supervised denoising methods, e.g. \cite{krull2019noise2void,batson2019noise2self,ulyanov2018deep} that avoid the need for explicit ground truth data. However, their performances are limited, particularly for low-light content since where, for example, edges are significantly modified in the presence of noise (examples can be seen in \cite{anantrasirichai:Contextual:2021}).


Conventional denoising methods involve performing processing at the pixel level using a weighted average or filtering of similar patches within the image, e.g. Non Local Means (NLM)\cite{buades2005non} and block-matching and 3D filtering denoising algorithm (BM3D) \cite{dabov2009bm3d}. Recently, methods based on deep learning have become the state of the art. DnCNN \cite{zhang2017beyond} was the first to outperform (in terms of PSNR) the conventional methods. DnCNN learns a mapping from the space of noisy images to the difference between the reference image and its degraded version. The intensive review of learning-based denoising algorithms can be found in  \cite{anantrasirichai:AI:2021}.  In the 2020 NTIRE denoising challenge (IEEE CVPR), using a Smartphone Image Denoising Dataset (SIDD) benchmark \cite{abdelhamed2018high}, the winners employed CNN variants, e.g. UNet \cite{ronneberger2015u} and ResNet \cite{he2016deep}, trained with $\ell_1$ loss. This denoising dataset however was generated using a range of lighting levels that are not representative of actual low-light conditions. In low light, noise originates from multiple sources and can be both additive and multiplicative as well as signal dependent. 

A significant limitation of the supervised denoising algorithms is that they require robust high quality and extensive ground truth data for training. Such data is difficult to acquire in practice. To overcome this, unsupervised and self-supervised learning techniques have been proposed \cite{krull2019noise2void,batson2019noise2self,ulyanov2018deep}, but they still cannot outperform the supervised learning approaches \cite{moreno2021evaluation} and generally face overfitting problems. To promote supervised learning approaches, in this paper, we present a new comprehensive dataset for low-light image denoising  -- BVI-LOWLIGHT \cite{bvilowlight}. Our dataset was acquired with varying ISO settings, offering a wide range of noise characteristics that are as well related to the topological properties of the image data. 

We further introduce a novel topological loss function which has been developed and characterised using this dataset. The novel loss combines the
local spatial (geometrical) image information and  global (topological) features of image patches. Its minimisation is equivalent to  $\ell_p$ norm minimisation together with convergence of the topological space of noisy images patches to the topological space of noise-free images patches.

This loss function is based on persistent homology, performing in the space of image patches, where topological invariants are calculated and represented in persistent diagrams. We compare the performance of the new loss function across popular denoising architectures, revealing that it supports adaption to  complex structures and is able to suppress common artifacts.



\section{Low-Light Image Denoising Dataset -- BVI-LOWLIGHT}
\label{sec:dataset}

\subsection{Statement of problem}
One of the shortcomings of most supervised denoising algorithms, particularly those based on deep learning, is the shortage of good-quality representative training datasets. In particular, the restoration of noisy low-light content is an ill-posed problem, so obtaining ground truth in most cases is difficult and time consuming.  Denoising algorithms are therefore frequently modelled and evaluated using synthetic noise, often with additive zero-mean Gaussian noise (AWGN). Such simple models do not accurately represent the noise present in natural images, particularly under low-light conditions, where photon noise dominates (sensor gain is implicitly coupled with noise amplification \cite{foi2008practical}). In order to evaluate the performance of denoising algorithms in these poor light conditions, we need either representative models or real noisy imagery with paired groundtruth. 

Existing state-of-the-art denoising datasets, such as RENOIR \cite{anaya2018renoir}, Darmstadt \cite{Plotz_2017_CVPR}, and SIDD \cite{abdelhamed2018high}, provide benchmarks for image denoising. However, these often lack content diversity. For example,   \cite{abdelhamed2018high} only has 10 scenes and \cite{Plotz_2017_CVPR} has only 6 ISO settings. They also have limited image resolution and bit depth, e.g.  $3684\times 2760$ pixels (8bit) \cite{anaya2018renoir} and $4032\times 3024$ pixels (8bit) \cite{abdelhamed2018high}. 
More importantly, the ISO values were set randomly for each scene in a range of 100-25600. This make it impossible to compare the impact of different ISO settings on the same content. Hence a comprehensive analysis of noise characteristics with respect to a sensor's sensitivity values cannot fully be performed on these datasets.

\subsection{BVI-LOWLIGHT dataset} 
To address the shortcomings  mentioned in the previous section, we have collected a new dataset (BVI-LOWLIGHT) \cite{bvilowlight}. Our dataset comprises 31800 14bit images of the size $4256\times 2848$ and $4948\times 3280$ in total, constituted from 20 scenes captured using 2 cameras (DSLR Nikon D7000 and mirrorless Sony A7SII). For each scene and camera type, 30 shots were taken with ISO settings ranging from 100 to 25600 (Nikon D7000) and 100 to 409600 (Sony A7SII). To provide consistent lighting environment with stable brightness and color temperature, we used non-flickering LED lights with fixed intensity and colour temperature values (our dataset involves different scenes with color temperature ranging from 2700K to 6500K and brightness levels varying between 2-4\%). We kept the focus point fixed on the object of interest, and we used different shutter speeds while keeping the aperture values fixed (6.3 Nikon D7000 for and 6.7 for Sony A7RII).
All images were captured using tripods and remote camera control library {\bf libgphoto2} to maximise stabilisation. The scenes contain a variety of content with a range of textures and colours.



\begin{figure*}
 
     \includegraphics[clip,width=\textwidth]{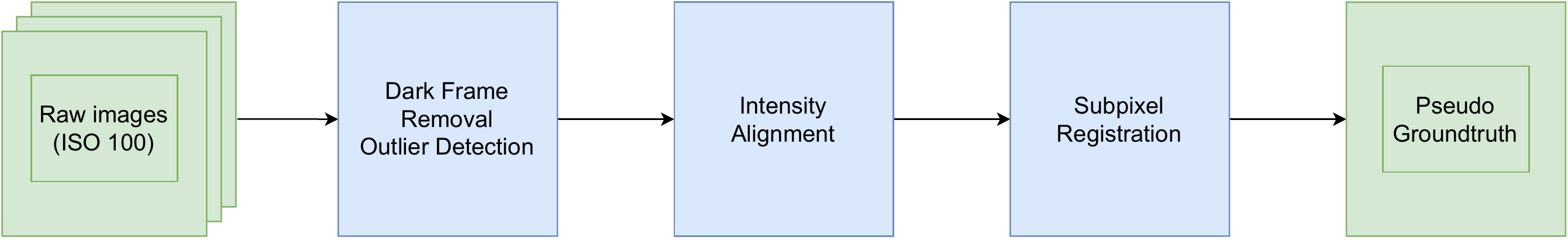}
     \caption{Dataset postprocessing workflow.}
    \label{fig:datapp}
\end{figure*}



\subsection{Pseudo Groundtruth Estimation}

The postprocessing pipeline, shown in Fig. \ref{fig:datapp}, is the key to create the reliable reference image (groundtruth) to use in training and evaluating. Here, we generate the pseudo groundtruth for each scene $s$ from  30 images $\mathcal{I}_i^s, i \in \{1,30\}$ with ISO value $100$, following the procedure in \cite{abdelhamed2018high}. 

First, we estimate the locations of oversaturated pixels on the sensor. These pixels can result in undesirable artifacts  -- pixel values with intensity values considerably higher than their neighbours. The oversatureated pixels can constitute up to 2.4\% of all pixels in the image \cite{zhang2004estimation}. To remove the defective pixels, we acquire a sequence of 200 dark images $\mathcal{I}^{dark}_{j}, {j \in \{1, ..., 200\}}$ for each camera. This is achieved by capturing frames in a dark studio environment, with no light sources, and with the lens tightly covered with a lens cap. For each camera, we calculate the standard deviation $\sigma$ and median $\mu$ of the distribution, detect all pixels $(x,y)\in\mathcal{I}$ for which
\begin{equation}
 \mathcal{I}{(x,y)} > \mu + \alpha\sigma,   
\end{equation}

\noindent where $\alpha$ is chosen so that $99.9\%$ of the pixels are within the confidence interval, and replace those pixels with their interpolated values $(\tilde{x}, \tilde{y})$, which are calculated by applying spatial median filter. We used median filtering as it is computationally cheap and is efficient in dealing with “salt-and-pepper” or “impulse” noise which is similar to hot pixels. We chose a size of 3$\times$3 to eliminate the outliers (hot pixels) while preserving the edges \cite{medianfilt2009}. 

Intensity alignment was also performed to reduce slight variations in lighting conditions and sensor characteristics. We manually picked and rejected a small number of the frames that had a considerable level of visible intensity shift. Mean intensity shift was performed on the remainder of the dataset through iteratively recalculating the mean value $m_a$ of the distribution of mean intensities $m_i$ across $i-th$ scene (see \cite{abdelhamed2018high} for more detail).

Although all images were acquired using stabilising tripods, some pixel-level displacements are possible within the image sequences. In \cite{guizar2008efficient} a pixel-level image registration technique based on the Fast Fourier Transform is performed to address this issue. However, for our dataset, we found that Euclidean rigid motion registration with the negative normalized cross correlation metric, linear interpolation, and gradient descent optimisation \cite{yaniv2018simpleitk} also provides the best results at a reasonable computational cost. 


The final step in pseudo groundtruth image estimation is averaging the set of pre-processed images from the previous steps for each scene. 

 Fig. \ref{fig:exampledata} shows some examples of our dataset (Complete set of scenes can be found in the supplemental material, see Fig. 1 and 2.). The inlets clearly demonstrate noise levels of different ISOs in different contents. Fig. \ref{fig:scenes_ssim_psnr} shows the image qualities in term of SSIM and PSNR at all ISO settings of Sony camera. The trend of most scenes (80\%) is monotonically decreasing.

\begin{figure*}
     \includegraphics[clip,width=\textwidth]{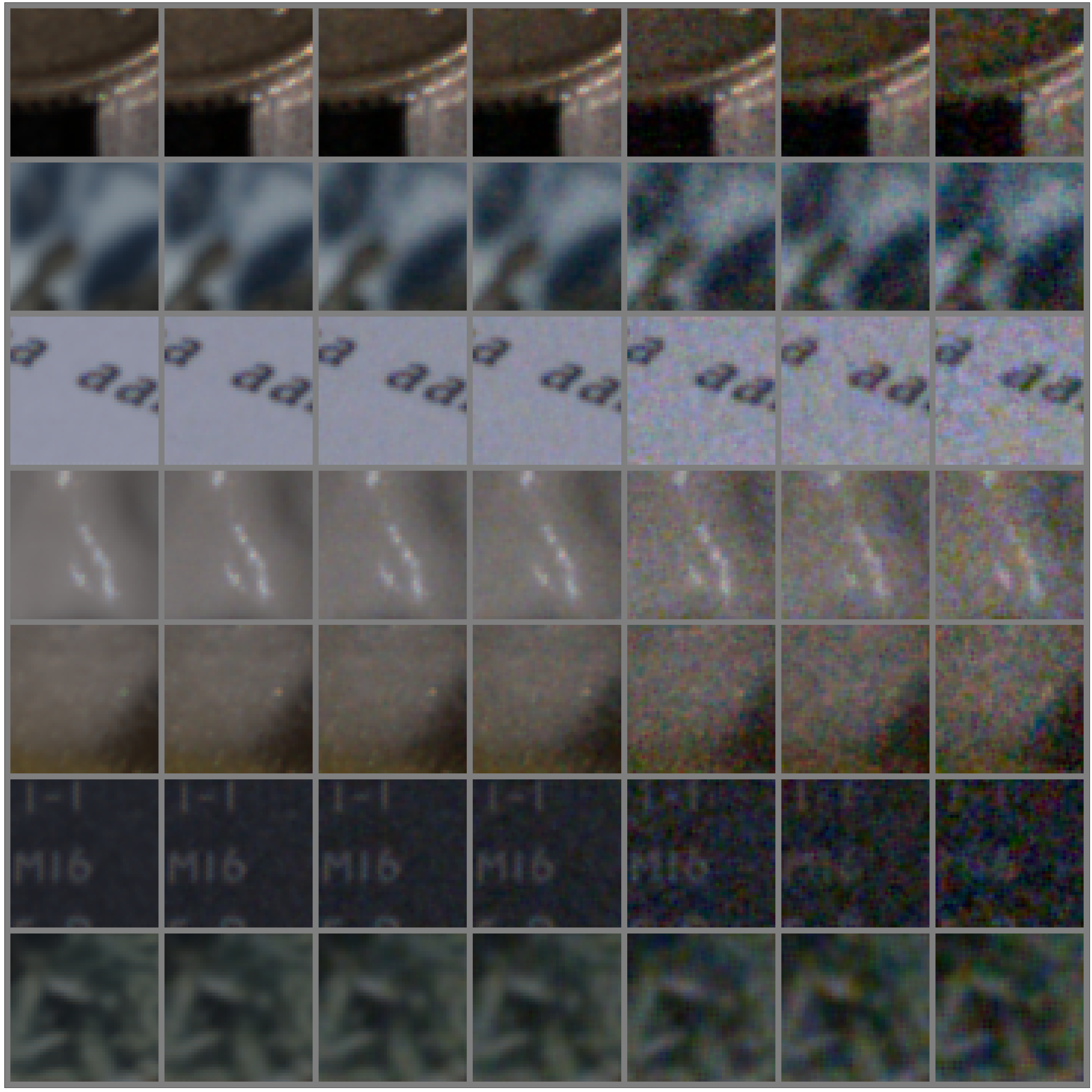}
     \caption{Examples of patches from BVI-LOWLIGHT dataset. First column: groundtruth images; second to the last columns:  noisy images with ISO = 1600, 6400, 12800, 64000, 102400, 160000 respectively.}
    \label{fig:exampledata}
\end{figure*}


\begin{figure}
  \subfloat[]{ 
    \includegraphics[width=.95\linewidth]{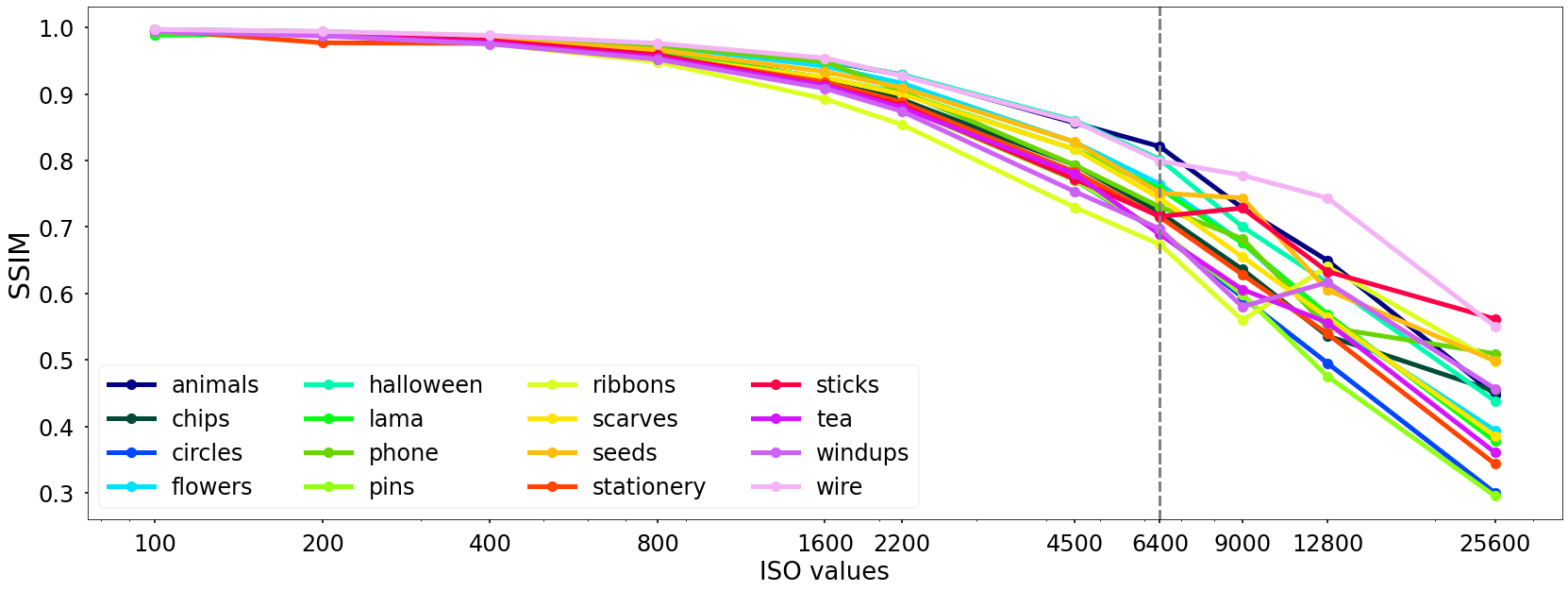}
    }


    \subfloat[]{
    \includegraphics[width=.95\linewidth]{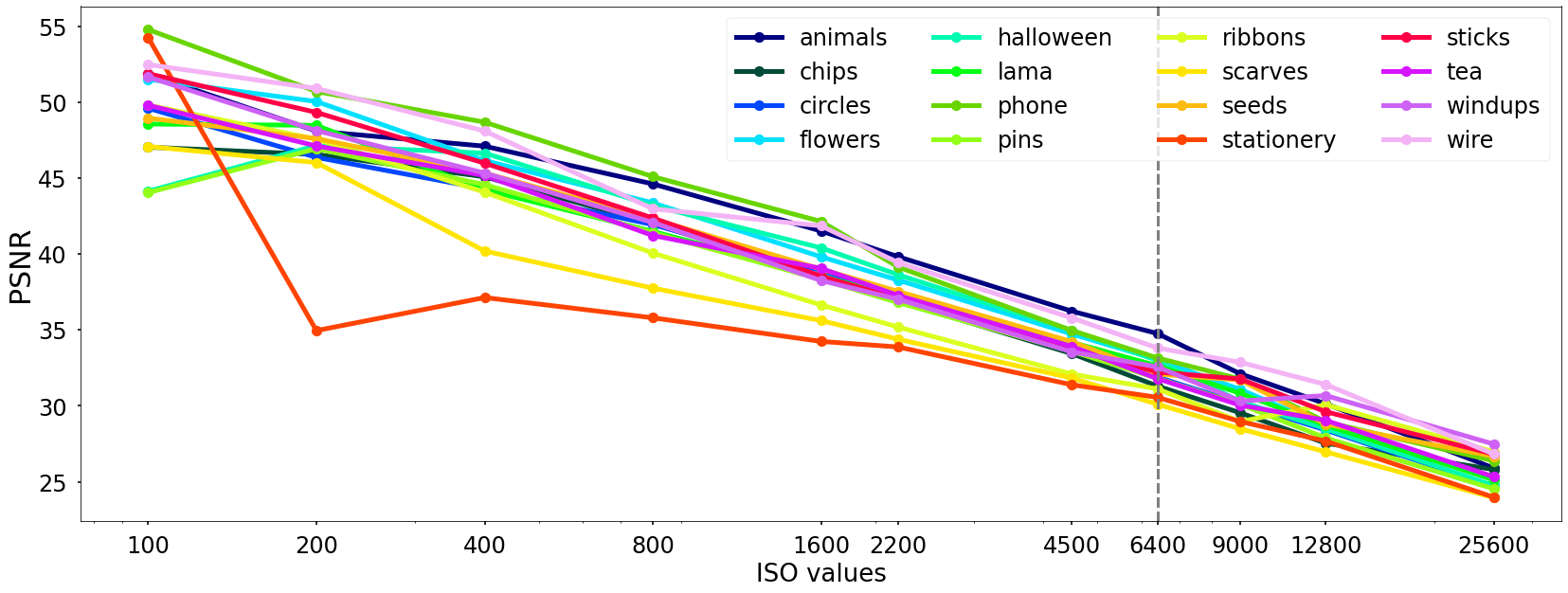}
  }
  \caption{ISO values vs SSIM (a) and PSNR (b). Dashed line indicates highest sensor-based camera ISO value above which the signal amplification is performed by camera software. The sensor may perform differently to the ISO standard above this value.}  
  \label{fig:scenes_ssim_psnr}
\end{figure}



\section{Topological Loss Function}
\subsection{Topological Data Analysis}

\label{sec:topoloss}


Tools from classical mathematical analysis and differential geometry, such as differentiable functions, metrics and integration, can provide knowledge about the distances and \textit{local structure} of data. However, they fail to provide information about global data structure. This data structure benefits denoising algorithms as it carries extra information that cannot be retrieved from the spatial image domain only.


In contrast, an approach based on topology can support analysis with respect to \textit{homeomorphisms} -- reversible continuous mappings, and hence tell us more about the \textit{global structure} of the data and its shape in the ambient space. Topological data analysis thus helps to utilise information about local and global structures and, together with statistical analysis, it provides information about the data that cannot be acquired using classical data analysis tools alone.


\subsection{Persistence Homologies}

Persistence homology has been an extensive area of research over the last two decades \cite{edelsbrunner2000topological,zomorodian2005computing,epstein2011topological} and has found application across biomedical image segmentation \cite{clough2019explicit,clough2020topological}, image analysis \cite{carlsson2008local}, network processes \cite{carstens2013persistent} and time series analysis \cite{seversky2016time,khasawneh2018chatter,gidea2018topological}. 

Persistent homology measures the features of data that persist across multiple scales and contains the information about the shape and structure of complex datasets.

Applying topological data analysis on discrete objects such as 3D point clouds or 2D images is not straightforward as topology generally studies continuous curves and manifolds.
Therefore, we need to introduce the notion of a complex and its filtration, which will underpin the calculation of persistence homologies and topological loss.



Let $V$ be a finite nonempty set, the elements of which are called vertices.
\textit{An (abstract) simplicial complex} on $V$ is a collection $\mathcal{C}$ of nonempty subsets of $V$ that satisfies the following conditions:

(i) $\forall v\in V$, the set $\{v\}$ lies in $\mathcal{C}$, 

(ii) $\forall \alpha\in \mathcal{C},$ and $\beta\subseteq \alpha$, $\beta$ is also an element of $\mathcal{C}$. Each $\alpha\in  \mathcal{C}$ is called \textit{a simplex}, each $\beta\subseteq \alpha$ is called \textit{a face}. The dimension of a simplex $\alpha\in \mathcal{C}$ is defined as $dim (\alpha) = |\alpha|-1$, the dimension of $\mathcal{C}$ is the highest dimension of constituent simplices.


Now let $\mathcal{C}$ be a simplicial complex. $\mathcal{C}'$ is called  \textit{a subcomplex of  $\mathcal{C}$}, if $\mathcal{C}'\subseteq\mathcal{C}$ and $\mathcal{C}'$ is a simplicial complex itself.


\textit{A (n-)filtration} of a simplicial complex $\mathcal{C}$ is a nested  sequence of subcomplexes (subsets) of $\mathcal{C}$:

\begin{equation}
    \mathcal{FC}_0\subset \mathcal{FC}_1 \subset \mathcal{FC}_2 \cdots \subset \mathcal{FC}_n= \mathcal{C}
\end{equation}

One can consider sublevel set filtrations of $\mathcal{C}$ that are defined by the function $f:\mathcal{C} \rightarrow \mathbb{R}$. The filtration is then defined by its increasing parameter $\varepsilon$, with $\mathcal{C}_{ \varepsilon } = f^{-1} (- \infty, \varepsilon]$. The distance-based filtrations are based on the pairwise distance of the points in the pointcloud $X=\cup_{i\in \mathcal{A}} \{x_i\}, x_i \in \mathbb{R}^n $: for any pair of points $x_i, x'_i \in X$, they are marked as ‘connected’ in a subcomplex $\mathcal{C}_\varepsilon$ if $d(x_i, x'_i) \leq \varepsilon$. 
\begin{figure}
\centering
     \includegraphics[width=0.8\textwidth,trim={0 10 30 0},clip]{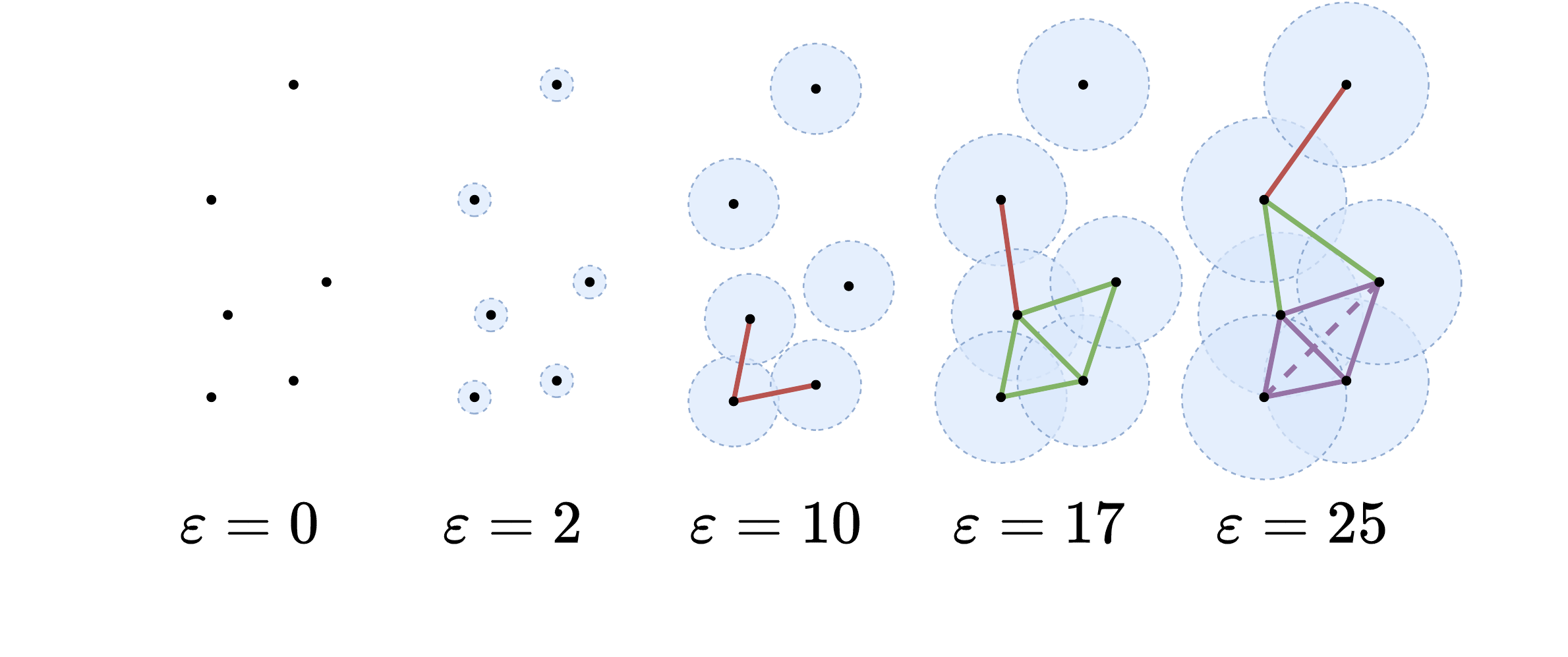}
     \caption{Distance-based filtration of Vietoris-Rips complex with subcomplexes $\mathcal{C}_{\varepsilon},$ where $\varepsilon\in\{0,2,10,17,25\}$. New simplices are formed with increasing radii $\varepsilon$ of the point neighbourhoods.}
    \label{fig:filt}
\end{figure}

We consider a simplicial complex with geometric realisation in $\mathbb{R}^n$, although the simplices need not be embedded in the space \cite{edelsbrunner2000topological}. Persistent homology studies an increasing filtration of simplicial complexes,  $\emptyset = \mathcal{C}_0 \subset \mathcal{C}_1 \subset ... \subset \mathcal{C}_n = \mathcal{C}$. 

We will not dive into the definitions of homology groups, one can find the comprehensive theoretical explanation in \cite{edelsbrunner2000topological}. Intuitively, homology classes $\mathbb{H}_k(\mathcal{C})$  \cite{edelsbrunner2000topological} of a complex are topological invariants that can be loosely interpreted as the number of connected components in $\mathcal{C}$ (for dimension $k=0$) or the number of holes (for dimension $k=1$).

The purpose of the filtrations is to capture the occurence and disappearance of topological features: the homology classes can appear and disappear with changing the parameter $\varepsilon$ of the filtration $\mathcal{C}_\varepsilon$ (see Fig. \ref{fig:filt} for Vietoris-Rips filtration). The filtrations of the complexes help to bring together local (distances) and global (shape) properties of the data.

A persistent diagram of a filtration $f$ on a complex $\mathcal{C}$ is a function
\begin{equation}
PD_k:(\mathcal{C},f)\to \{b_i, d_i\}_{i\in I_k},    
\end{equation}

\noindent that maps every $i^{th}$ $k$-dimensional topological feature into a pair $(b_i,d_i)\in \mathbb{R}^2\cup \{\infty\}$ where $b_i$ indicates the appearance (“birth”) of the topological feature when $\varepsilon=b_i$ and its disappearance (“death”) when $\alpha=d_i$. For our purposes we only consider $k=0$ that corresponds to the number of connected components in filtered subcomplexes of $\mathcal{C}$ and $k=1$ --- the number of 1-dimensional holes.



\subsection{Topological Loss Function}

A subset of contrast patches from natural images has an intrinsic topological structure of a Klein bottle \cite{lee2003nonlinear}, i.e. a two-dimensional non-orientable manifold \cite{hatcher2000algtop}. This can be explained by the fact that natural images contain sharper edges, and hence the corresponding patches have higher gradients \cite{adams2009nonlinear}. A study in \cite{carlsson2008local} confirms that the space of $3\times 3$ patches of natural images has useful topological properties. We therefore utilise the space of $3\times 3$ image patches to build a new loss function. 


To create the image patch space, we performed the following procedure \cite{adams2009nonlinear}:

\begin{enumerate}

    \item We split each image $\mathcal{I}$ into the set of $3\times3$ patches $\{p_i\}_{i\in\mathcal I}$. 
    \item For each patch $p$ we calculate its contrast norm (D-norm) as follows:  
    \begin{equation}
    ||p||_D = \sum_{i,j} \sqrt{(p_i-p_j)^2}, 
    \end{equation}
    where $p_i$ and $p_j$ are adjacent pixels in the patch $p$.
        
    \item We select the top $t$ highest contrast patches. These carry the important structural information \cite{adams2009nonlinear}. For our experiment we set $t$ to $0.2$, so that we obtain $20\%$ of the most contrast patches. 
    \item The obtained patch space is normalised by subtracting the mean from the coordinate of a patch vector and dividing it by its norm, hence mapping the patch space onto a $9$-dimensional sphere $\mathbb{S}^9$.
    
    \item In order to derive the underlying topology of the image patch space, we need to remove ``outlier" points and represent the patch space by its densest elements: the $k$-density is calculated as the distance from the chosen point to its $k^{th}$ nearest neighbour. 
    For further computational efficiency, we randomly sample $n$ elements from this set of densest patches (e.g. for training the denoising model with $256\times 256$ cropped image, we sample 300 elements with $k=30$ to compute topological loss). 
    \end{enumerate}

\begin{figure}[ht]
    \centering
    \subfloat[]{ 
        \includegraphics[width=0.5\linewidth,trim={25 16 25 25 },clip]{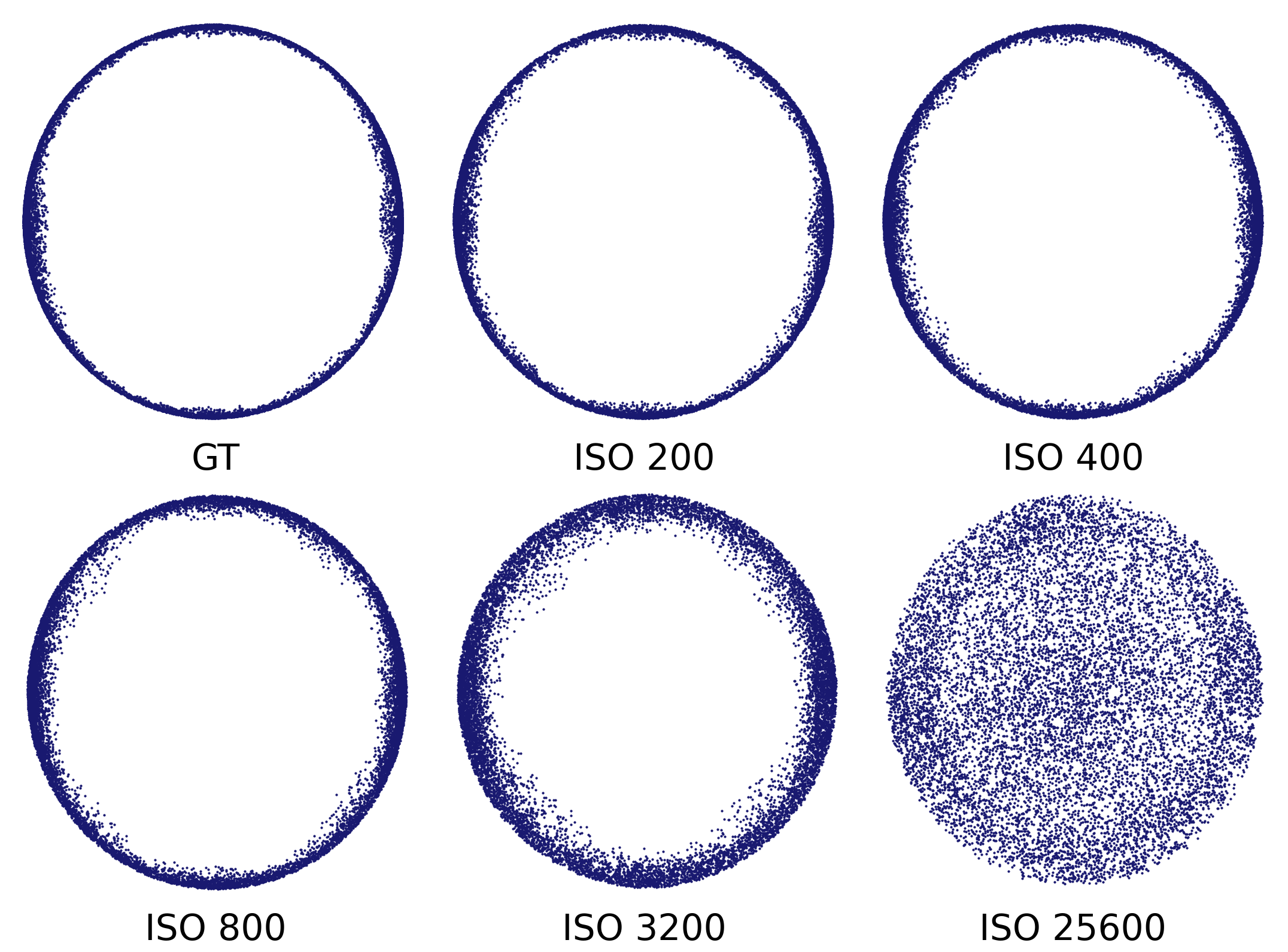}
}\qquad
      \subfloat[]{ 
        \includegraphics[width=0.35\linewidth,trim={15 5 15 15},clip]{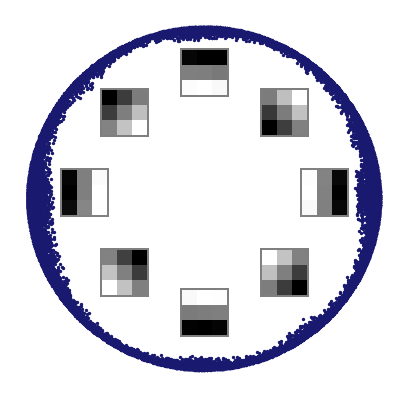}
      }
        \label{fig:patches}
       \caption{(a) PCA projections of the patch space $\mathcal{P}_{(30,300)}$ sampled from BVI-LOWLIGHT dataset, at different ISO levels. The distribution of patches tends to be more dense across the principal patches for lower ISOs and scattered for higher ISOs (noisy). (b) Principal $3\times3$ patches of the PCA projection of the patch space $\mathcal{P}_{(30,300)}$ onto 2-dimensional plane for the image with ISO 100.}

    \label{fig:pp}
\end{figure}

We denote by $\mathcal{P}_{(k,n)}$ the resulting sampled subset of $3\times3$ patches (represented as 9-vectors). As the computational complexity of persistent diagram calculation is cubic with respect to the number of data points \cite{otter2017roadmap}, we consider $3\times3$ patches from $\mathcal{P}_{(30,300)}$ to achieve a reasonable compromise in terms of computation time. PCA projections of the patch spaces illustrating the difference in distributions of images with different ISO levels from the dataset are shown in Fig. \ref{fig:pp}.

\begin{figure}
    \centering
    \includegraphics[width=.95\linewidth]{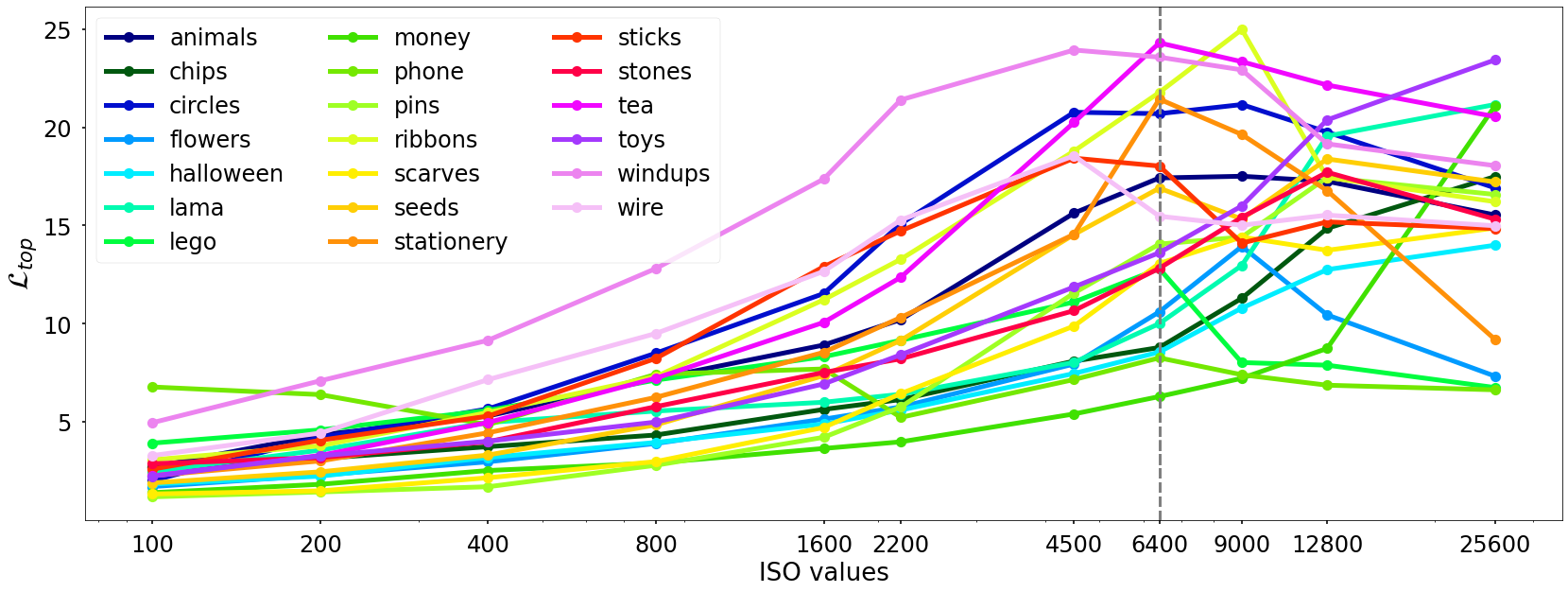}
  \caption{ISO values vs values of topological loss $\mathcal{L}_{top}(30,300)$. Dashed line indicates highest sensor-based camera ISO value above which the signal amplification is performed by camera software. The sensor may perform differently to the ISO standard above this value.}  
  \label{fig:top_loss}
\end{figure}

Having two subsets $\mathcal{P}^{N}_{(k,n)}$ and $\mathcal{P}^{C}_{(k,n)}$  of noisy $\mathcal{I}^{N}$ and clean $\mathcal{I}^{C}$ images, respectively, 
we now define topological loss term as follows:
\begin{equation}
\label{eq:ltop}
\mathcal{L}_{top}(\mathcal{I}^{N}, \mathcal{I}^{C}) = 
W_p(PD(\mathcal{P}^{N}_{(k,n)}),PD(\mathcal{P}^{C}_{(k,n)})),
\end{equation}
\noindent where 
\begin{equation}
W_p(PD_1,PD_2)= \left( \inf_{x\in P_1, y\in P_2} \mathbb{E} \big[||x-y||_2 \big] \right)^{1/p}
\end{equation}
\noindent is the $p$-Wasserstein distance between persistence two diagrams $P_1$ and $P_2$ \cite{olkin1982distance}. Figure \ref{fig:top_loss} shows values of $\mathcal{L}_{top}^{(30,300)}$ calculated the for variable  dataset.

For the denoising task, we define our combined \textit{topological loss function} as a scaled combination of a loss $\mathcal{L}_{top}$, defined earlier in [6], and a base loss $\mathcal{L}_{base}$:
\begin{equation}
    \mathcal{L}_{comb} = \alpha \mathcal{L}_{top}  + \beta \mathcal{L}_{base},
\end{equation}

 \noindent where $\mathcal{L}_{base}$ is $\ell_1$ or $\ell_2$ loss and $\alpha, \beta$ are scaling coefficients (in our case, $\alpha=0.93, \beta = 0.07$ were chosen through parameter search).
 $\mathcal{L}_{base}$ is conventionally used for image denoising and operates in image spatial domain while $\mathcal{L}_{top}$ provides convergence in the space of topological descriptors (as shown in persistent diagrams). Therefore, our task can be formulated as a minimisation problem, where we search for an optimal parameter set $\bm{\theta} = \{\theta_i\}_{i\in\mathcal{I}}$ for a denoising model $f_{\bm{\theta}}$:

\begin{equation*}
\hat{\bm{\theta}}=\arg \min_{\bm{\theta}} \mathcal{L}_{comb}(f_{\bm{\theta}}(\mathcal{I}^{N})).
\end{equation*}

\begin{figure}
\centering
     \includegraphics[width=0.96\textwidth]{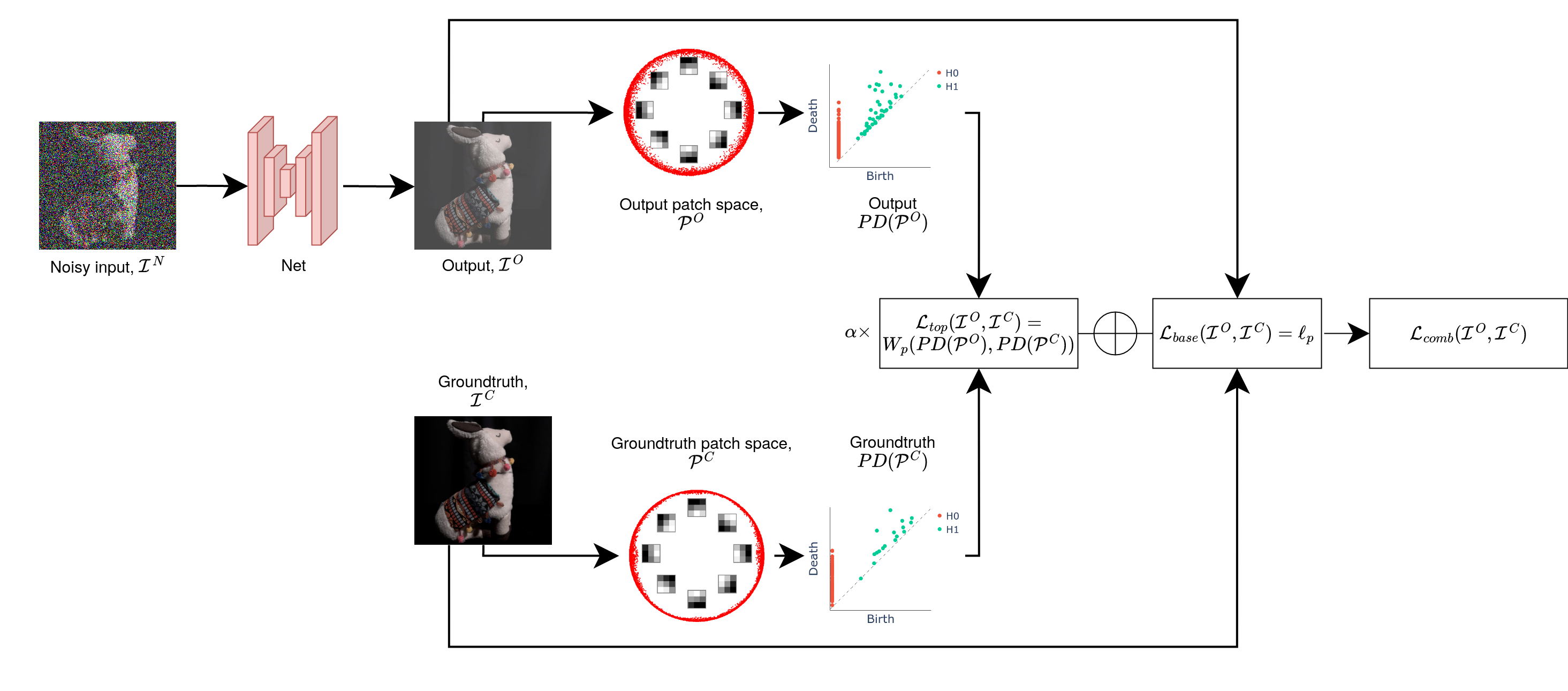}
     Combined topological loss \caption{$\mathcal{L}_{comb}$ calculation for a pair of images. First, the pair of noisy image $\mathcal{I}_N$ and output from the network $\mathcal{I}_O$ are translated into their corresponding patch spaces $\mathcal{P}_N$ and $\mathcal{P}_O$. The topological component $\mathcal{L}_{top}$ is calculated as Wasserstein distance between the persistent diagrams $PD(\mathcal{P}_N)$ and $PD(\mathcal{P}_O)$. We also calculate  $\mathcal{L}_{base}$ ($l_p$ loss with $p=1$ or $2$) to retain image spatial information. The resulting combined topological loss  $\mathcal{L}_{comb}$ is calculated as weighted sum of  $\mathcal{L}_{top}$ and  $\mathcal{L}_{base}$.}
    \label{fig:scheme}
\end{figure}

The combined loss calculation pipeline is shown on figure \ref{fig:scheme}.

\section{Experiments and discussion}

We used our proposed topological loss function constructed in for denoising low-light images with three state-of-the-art architectures: i) a residual-based denoising convolutional neural network (DnCNN) introduced in  \cite{zhang2017beyond} by Zhang et al., widely-used method for benchmarking deep
image denoisers, ii) UNet, an  encoder-decoder style fully convolutional network  \cite{ronneberger2015u}, having been used as a backbone of image denoisers, iii) CGAN, conditional generative adversarial network, specifically Pix2Pix \cite{isola2017image} that was originally introduced for image translation; iv) RIDNet, a single-stage blind real image denoising network based on feature attention mechanism \cite{anwar2019ridnet}.

We noticed that the standard UNet architecture suffers from checkerboard artifacts caused by the use of transpose convolutions in its upsampling layers \cite{shi2016deconvolution} \cite{shi2016real}. Using pixel shuffle in the upsampling layer instead of transpose convolutions provides a significant gain of $\approx 4$dB for the UNet architecture trained on our dataset. For DnCNN architecture we used the standard 17 convolutional layers followed by batch normalization and ReLU activation. For Pix2Pix architecture we used UNet as a generator network, identical to chosen UNet architecture. The model architectures are provided in the supplemental material (Fig. 3-6).

\begin{figure}[H]
\centering
     \includegraphics[width=0.8\textwidth]{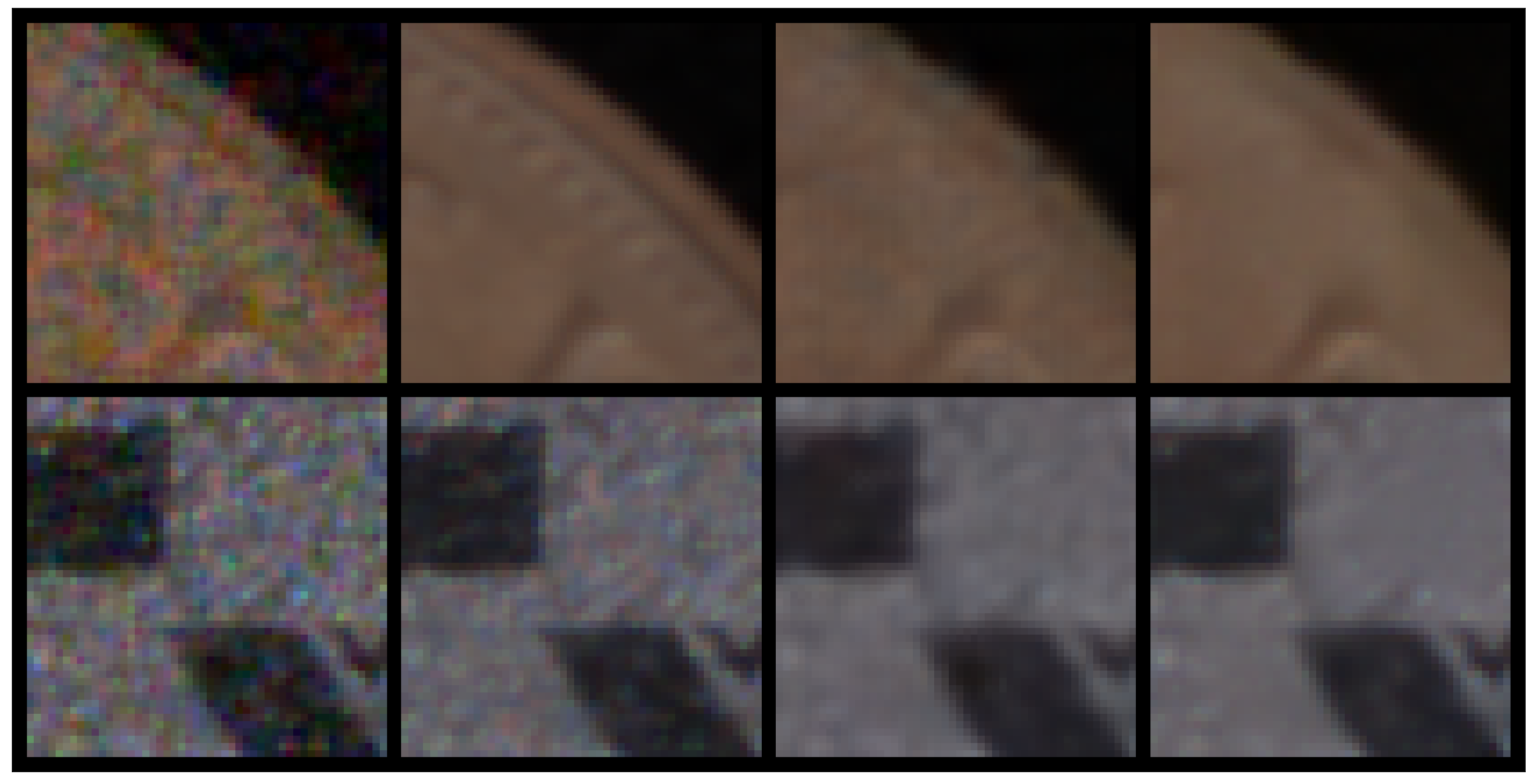}
     \caption{Left to right: a patch  from a noisy image; groundtruth; result from $\mathrm{Pix2Pix}$ with $\mathrm{UNet}$ generator; result from plain $\mathrm{UNet}$ model. Combined topological loss was used for training all models. Generative model produces textures in some examples, while plain model tends to flatten them.}
    \label{fig:ganex}
\end{figure}

\begin{figure}[H]
   \centering
   
    \includegraphics[height=0.7\textwidth]{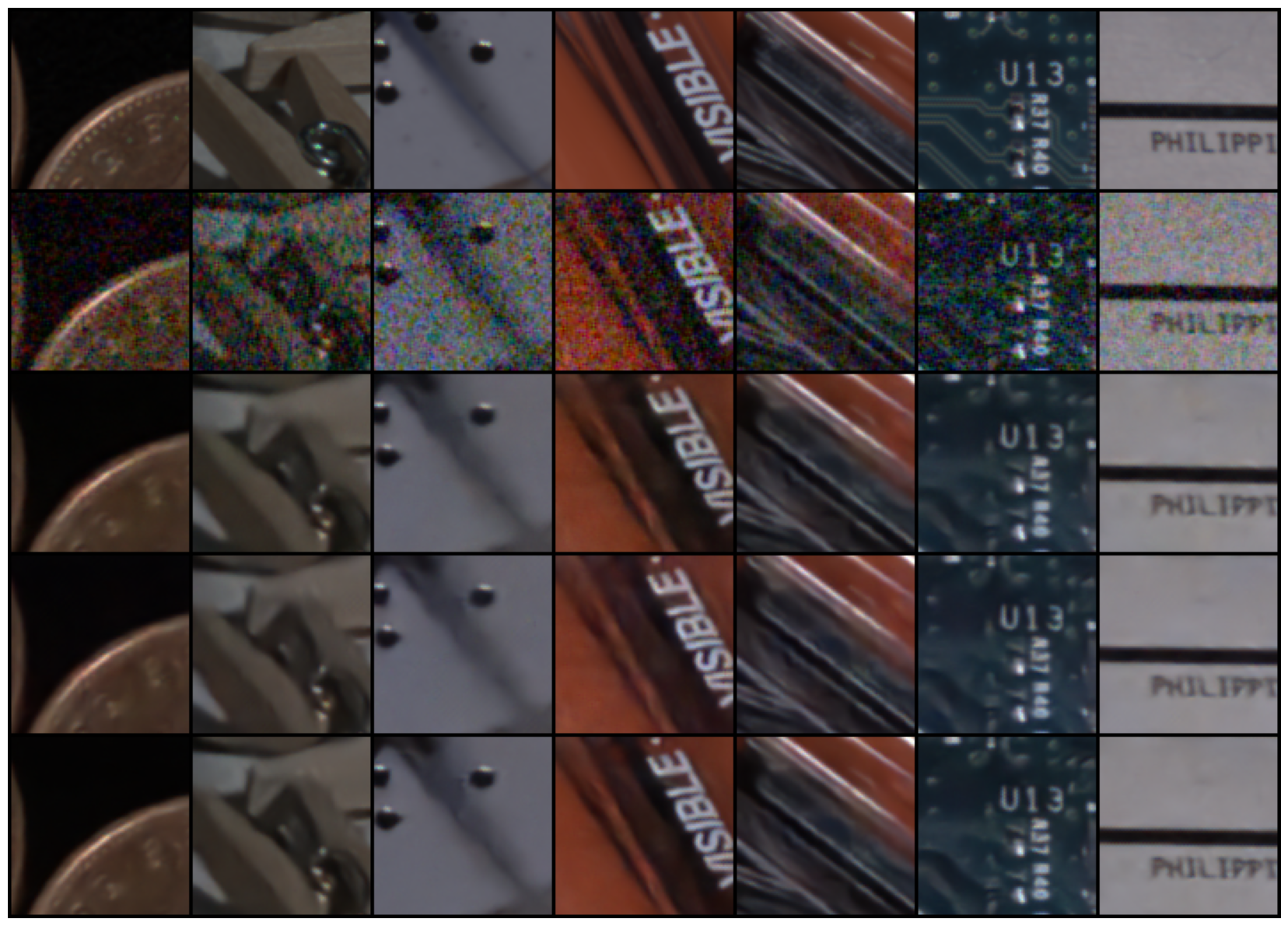}

    
    \caption{Subjective results-I. \textit{Top row}: Estimated ground truth.  \textit{Second row}: Noisy image patches with ISO varying from 80000 to 400000. \textit{Third row}: Outputs from RIDNet trained with $\ell_1$ only.
    \textit{Fourth row}: Outputs from RIDNet trained with $\mathcal{L}_{vgg}+\ell_1$.
    \textit{Bottom row}: Outputs from RIDNet trained with $\mathcal{L}_{topo}+\ell_1$. Note the slight colour variation and the enhanced texture (lines and contrast in fine details)}
     \label{fig:resultsridnet}
\end{figure}

\begin{figure}[H]
   \centering
   
    \includegraphics[height=0.6\textwidth]{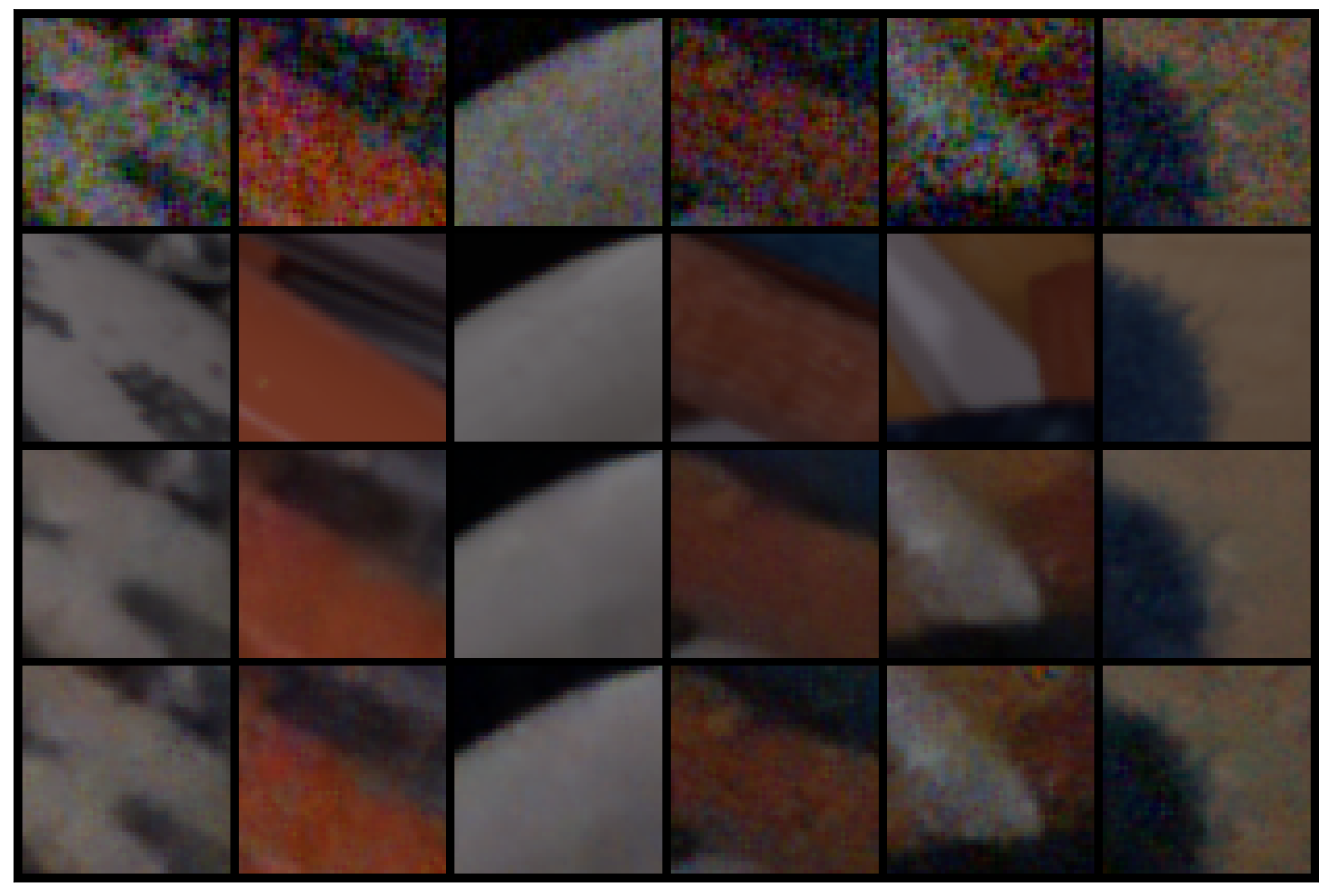}

    
    \caption{Subjective results-II. \textit{Top row}: Noisy image patches with ISO varying from 160000 to 400000.  \textit{Second row}: Estimated ground truth. \textit{Third row}: Outputs from DnCNN trained with topology loss combined with $\ell_1$. \textit{Bottom row}: Outputs from DnCNN trained with $\ell_1$ only. Note the artifacts present when the persistent homology loss are not used.}
     \label{fig:results}
\end{figure}

\begin{figure}[H]
\centering
     \includegraphics[width=0.9\textwidth]{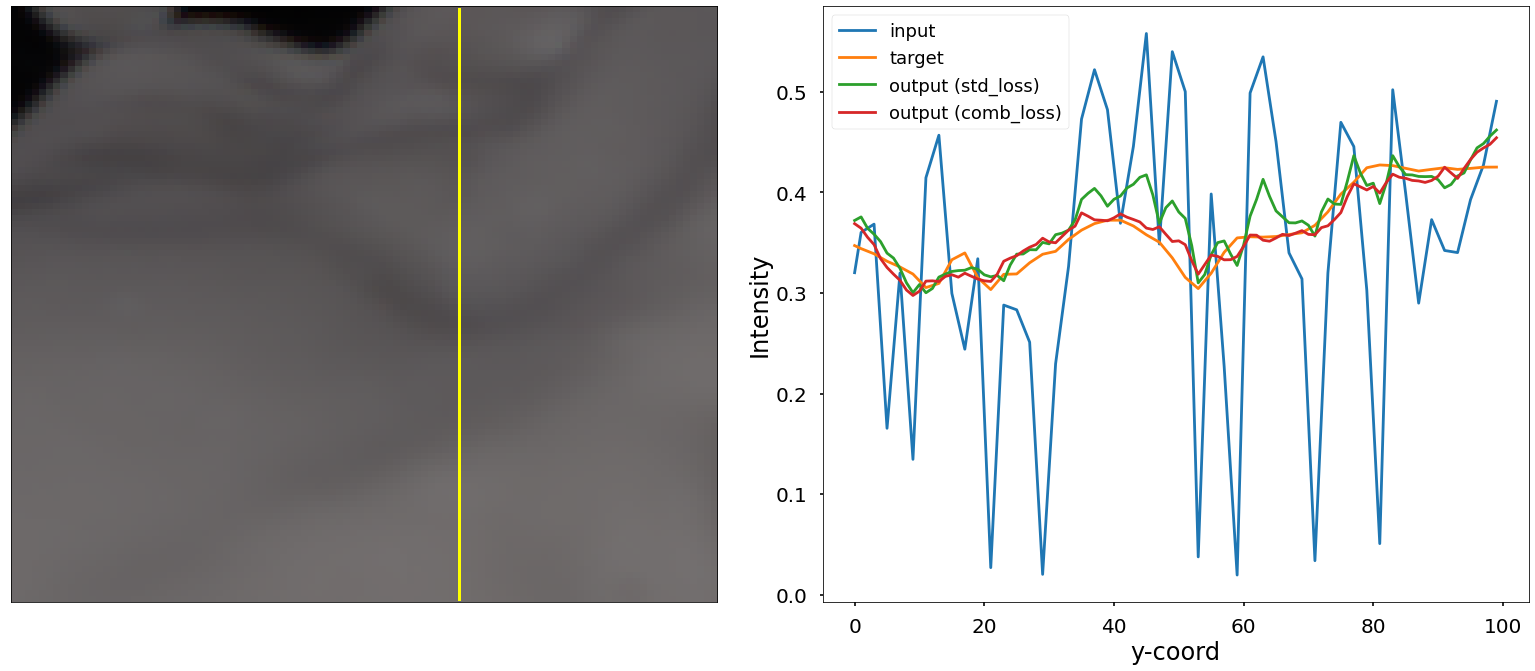}
     \caption{Left: Groundtruth image with selected pixel column (yellow). Right: Pixel value gradient profile of the selected column, calculated for input noisy image (blue), target image (orange) and outputs from DnCNN trained with topological loss component (red) and DnCNN trained with $\ell_1$ only (green). The gradients of the red line are smoother and closer to the target. This indicates that the topological loss component is pushing the noisy image patch space to that of the clean image.}
    \label{fig:profile}
\end{figure}

We trained each of the networks for 40 epochs with learning rate $lr = $ 0.0001 using Adam optimiser. The training dataset consisted of 21600 patches of the size $256\times256$ with ISO values varying from 200 to 409600. The patches are randomly drawn from the images covering the full range of scenes in the dataset. The resulting metric values (PSNR and SSIM) for the test set can be found in Tables \ref{table:metrics:psnr} and  \ref{table:metrics:ssim} and the subjective results are shown in Figures \ref{fig:ganex}, \ref{fig:resultsridnet}, \ref{fig:results}, \ref{fig:profile}. We also trained the models using VGG loss ($\mathcal{L}_{top}$) as it proved its effectiveness in image transformation tasks \cite{vggloss2016}. As VGG loss term is also used in combination with standard losses, we use both $\ell_1$ and $\ell_2$ as base loss functions for it.

The inclusion of the topology loss component $\mathcal{L}_{top}$ resulted in increased PSNR and SSIM values over using conventional losses for all but one model. However, for some architectures combining topology loss with $\ell_2$ was insignificant, especially comparing to VGG, which relies on low-level image features.

Although architectures with UNet used as a generator of CGAN provide lower objective metric scores they are in some cases perceptually more pleasant due to the ability of GANs to generate textures learned from ground truth images distribution (see Fig. \ref{fig:ganex} ).

\begin{table}[ht]

\centering
\begin{tabular}{lllllll}
\hline
\multicolumn{1}{|l|}{}              & \multicolumn{1}{l|}{${\ell_1}$}    & \multicolumn{1}{l|}{${\ell_1}+\mathcal{L}_{vgg}$} & \multicolumn{1}{l|}{${\ell_1}+\mathcal{L}_{top}$}                       & \multicolumn{1}{l|}{${\ell_2}$}    & \multicolumn{1}{l|}{${\ell_2}+\mathcal{L}_{vgg}$} & \multicolumn{1}{l|}{${\ell_2}+\mathcal{L}_{top}$}                       \\ \hline

\multicolumn{1}{|l|}{$\mathrm{DnCNN}$}   & \multicolumn{1}{r|}{38.01} & \multicolumn{1}{r|}{39.35}  & \multicolumn{1}{r|}{\cellcolor[HTML]{C5FDB3}39.42} & \multicolumn{1}{r|}{37.50} & \multicolumn{1}{r|}{38.63}  & \multicolumn{1}{r|}{39.20}        \\ \hline     
\multicolumn{1}{|l|}{$\mathrm{UNet}$}   & \multicolumn{1}{r|}{40.23} & \multicolumn{1}{r|}{40.88}  & \multicolumn{1}{r|}{40.77} & \multicolumn{1}{r|}{38.79} & \multicolumn{1}{r|}{41.31}  & \multicolumn{1}{r|}{\cellcolor[HTML]{C5FDB3}41.95}      \\ \hline   
\multicolumn{1}{|l|}{$\mathrm{Pix2Pix}_\mathrm{UNet}$}  & \multicolumn{1}{r|}{39.01} & \multicolumn{1}{r|}{38.42}  & \multicolumn{1}{r|}{39.35}                         & \multicolumn{1}{r|}{37.50} & \multicolumn{1}{r|}{38.63}  & \multicolumn{1}{r|}{\cellcolor[HTML]{C5FDB3}40.20}                        \\ \hline
\multicolumn{1}{|l|}{$\mathrm{RIDNet}$}           & \multicolumn{1}{r|}{44.15} & \multicolumn{1}{r|}{44.89}  & \multicolumn{1}{r|}{\cellcolor[HTML]{C5FDB3}45.75} & \multicolumn{1}{r|}{43.50} & \multicolumn{1}{r|}{44.00}  & \multicolumn{1}{r|}{45.11}                         \\ \hline
                                  
\end{tabular}
\caption{PSNR values for i) DnCNN, ii) UNet, iii) Pix2Pix with UNet generator and iv) RIDNet trained with ${\ell_1}$ and ${\ell_2}$ alone, and combinations of ${\ell_1}$ and ${\ell_2}$ with persistence-based loss $\mathcal{L}_{top}$ and VGG loss $\mathcal{L}_{vgg}$. }
\label{table:metrics:psnr}
\end{table}

\begin{table}[ht]

\centering
\begin{tabular}{|l|r|r|r|r|r|r|}
\hline
                 \multicolumn{1}{|l|}{}              & \multicolumn{1}{l|}{${\ell_1}$}    & \multicolumn{1}{l|}{${\ell_1}+\mathcal{L}_{vgg}$} & \multicolumn{1}{l|}{${\ell_1}+\mathcal{L}_{top}$}                       & \multicolumn{1}{l|}{${\ell_2}$}    & \multicolumn{1}{l|}{${\ell_2}+\mathcal{L}_{vgg}$} & \multicolumn{1}{l|}{${\ell_2}+\mathcal{L}_{top}$}      \\ \hline
$\mathrm{DnCNN}$             & 0.880                   & 0.898                       & \cellcolor[HTML]{C5FDB3}0.928 & 0.853                   & 0.894                       & 0.914                         \\ \hline
$\mathrm{UNet}$      & 0.946                   & 0.952                       & \cellcolor[HTML]{C5FDB3}0.957 & 0.953                   & 0.953                       & 0.957                        \\ \hline
$\mathrm{Pix2Pix}_\mathrm{UNet}$         & 0.880                   & 0.899                       & 0.915                         & 0.853                   & 0.895                       & \cellcolor[HTML]{C5FDB3}0.928 \\ \hline
{$\mathrm{RIDNet}$}              & 0.945                   & 0.940                       & 0.957                         & 0.956                   & 0.956                       & \cellcolor[HTML]{C5FDB3}0.963 \\ \hline
\end{tabular}

\caption{SSIM values for i) DnCNN, ii) UNet, iii) Pix2Pix with UNet generator and iv) RIDNet trained with ${\ell_1}$ and ${\ell_2}$ alone, and combinations of ${\ell_1}$ and ${\ell_2}$ with persistence-based loss $\mathcal{L}_{top}$ and VGG loss $\mathcal{L}_{vgg}$. }

\label{table:metrics:ssim}

\end{table}

Regarding the limitations of our topological loss function, further efforts are required to improve its computational efficiency. Computational speed remains an issue for real-time applications: the computational time for combined topological loss $\mathcal{L}_{comb}$ for our setup is 4$\times$ time for $\mathcal{L}_{vgg}$ loss and 8$\times$ time for $\ell_1$ or $\ell_2$ loss. This also poses challenges when searching for optimal parameters.




\section{Conclusions}
We have introduced a novel topological loss based on the calculation of persisent homologies in the space of image patches and applied this to the problem of denoising low light imagery. Our results show that this $\mathcal{L}_{top}$ loss term helps the performance of denoising architectures compared to training with traditional loss functions. Subjective results show that the use of topology-inspired loss function helps to reduce artifacts while still preserving edges and contrast.  This can be explained by the fact that the $\mathcal{L}_{top}$ term of combined topological loss is encouraging the models to push the patch space of the noisy image to be topologically similar to that of noiseless natural images, which produces images that are not only noise-free but also visually and structurally similar to natural images.
Further improvement can be made by tuning the parameters of $\mathcal{L}_{top}$ (e.g. $k$ in $k$-nearest neighbours, dimension of persistent diagrams) or incorporating other tools of topological data analysis such as persistent landscapes or total persistence \cite{otter2017roadmap}, as well as optimising the computations.



\bibliographystyle{elsarticle-num} 
\bibliography{biblio}

\end{document}


\section*{Supplemental material}

\title{}

\begin{figure*}[h!]

\centering
    \includegraphics[width=\linewidth]{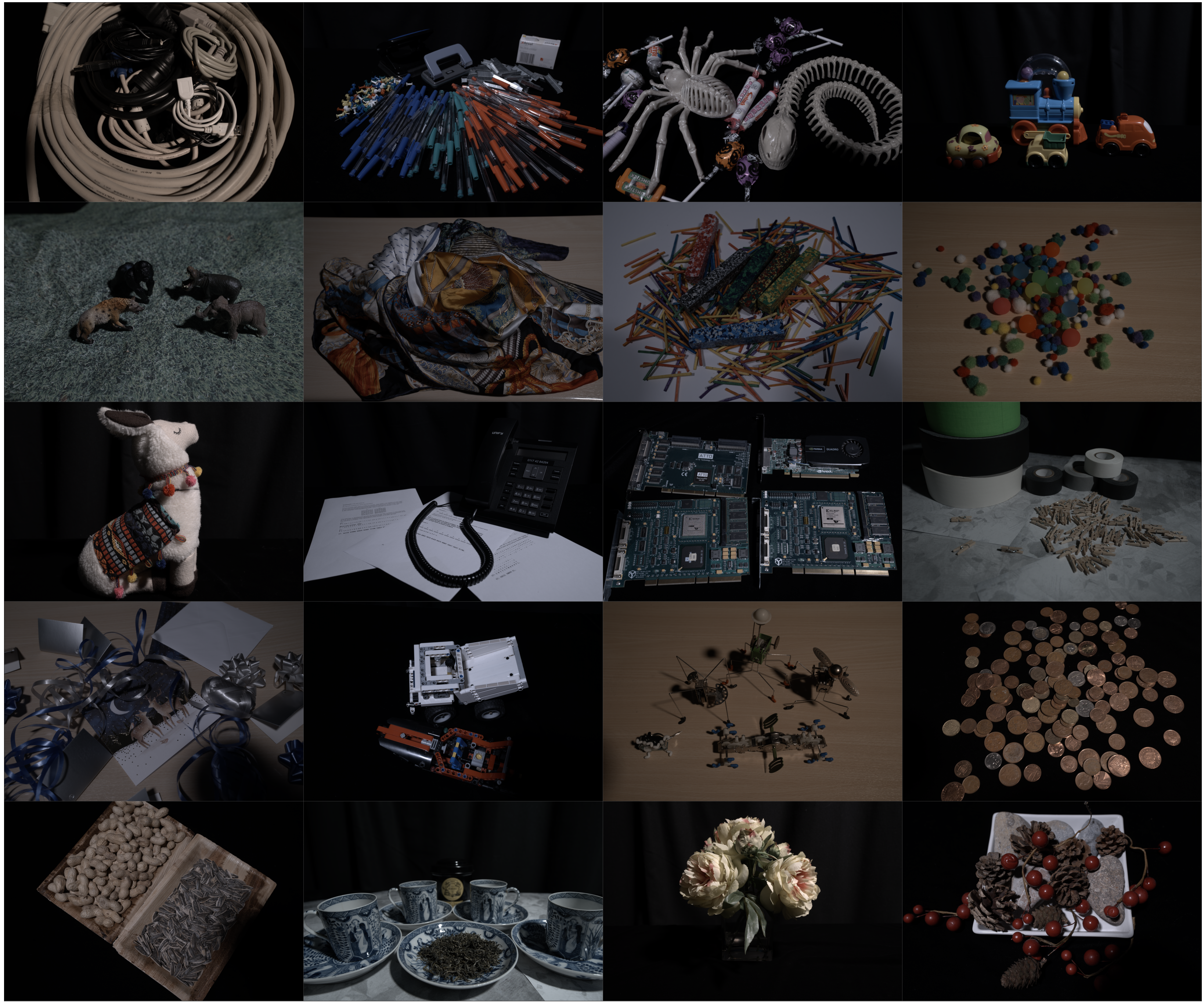}
     \caption{20 scenes from BVI-LOWLIGHT dataset. The total dataset consists 31800  images captured using 2 cameras: Nikon D7000 (4948x3280) and Sony A7SII (4256x2848). 20 scenes captured in low light conditions provide with a wide variety of textures and content.}
\end{figure*}

\begin{figure*}[ht]

\centering
    \includegraphics[width=.8\linewidth]{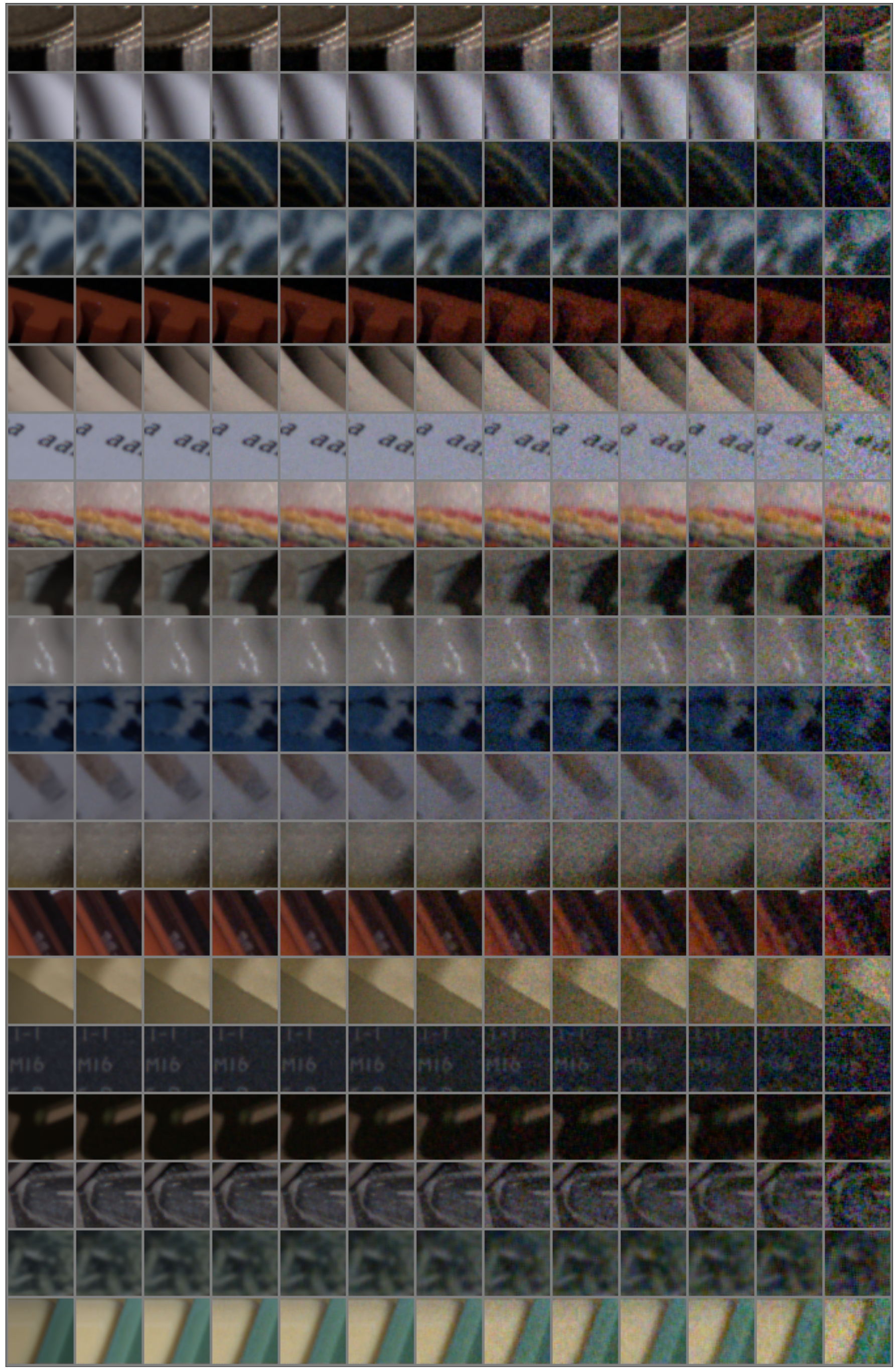}
     \caption{BVI-LOWLIGHT dataset sample patches. For each scene ISO values range from 100 (left) to 409600 (right).}
\end{figure*}
\begin{figure*}[ht]

\centering
    \includegraphics[width=.6\linewidth]{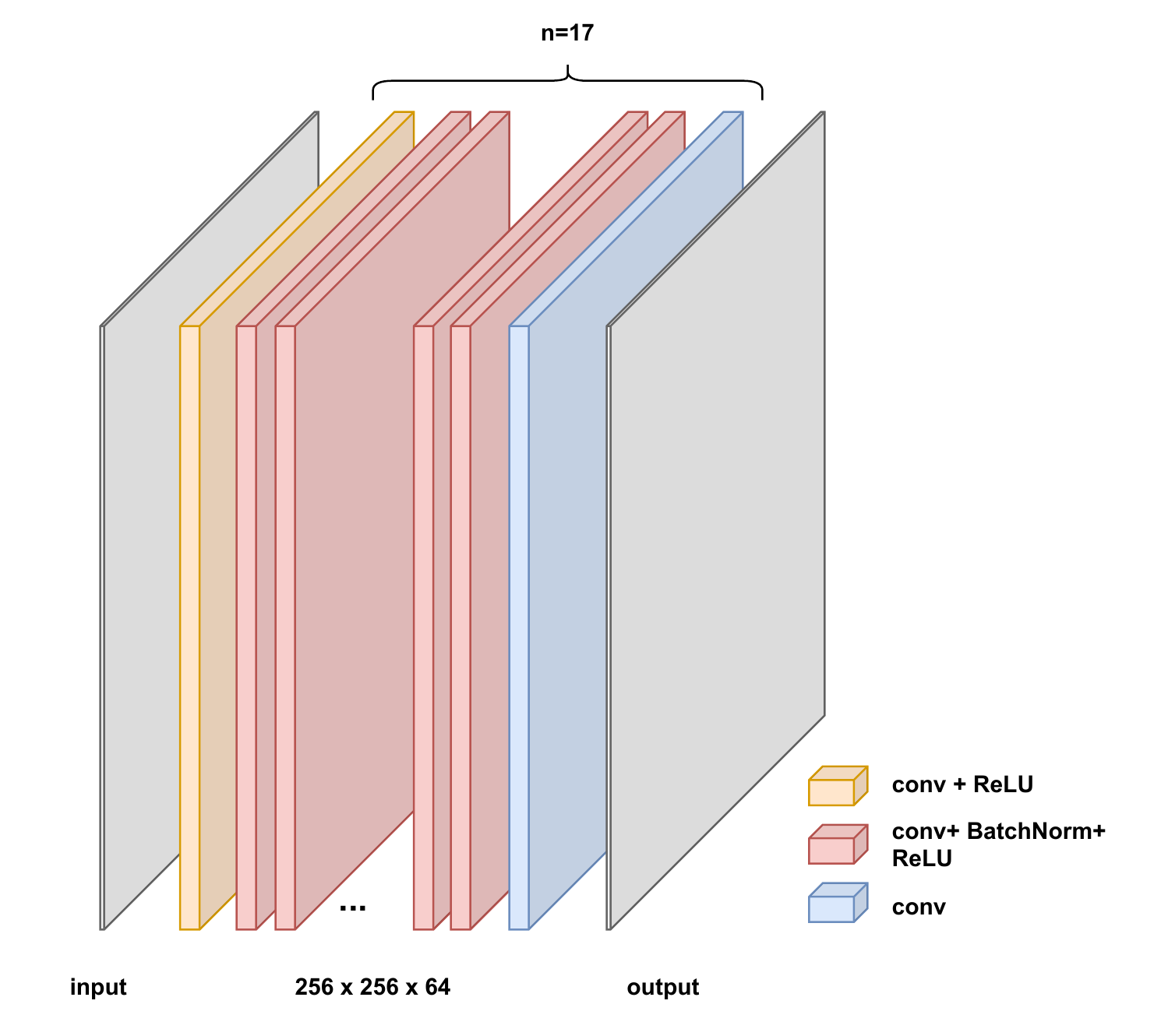}
     \caption{DnCNN architecture. We used 17 convolutional blocks, each of which consists of convolution, batch normalization and ReLU layers (apart from the final layers, where activations are skipped).}
\end{figure*}

\begin{figure*}[ht]
\centering
        \includegraphics[width=.7\linewidth]{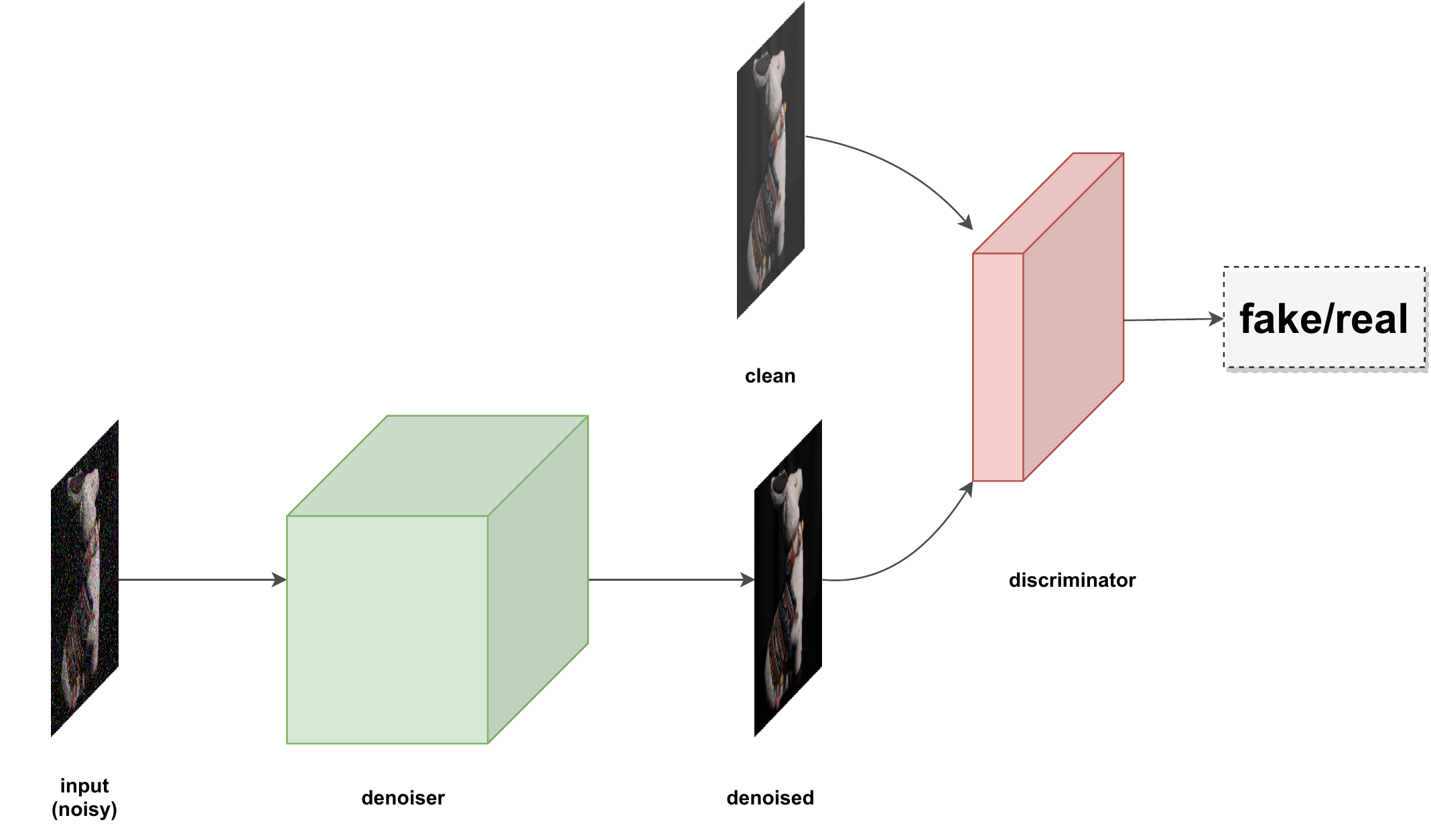}
         \caption{Pix2Pix diagram. The generator given a noisy image as input and outputs the denoised version of the image. The transformed image is then given to the discriminator which must determine whether the image is a (real) ``clean'' image or the denoised version produced by the generator (fake). Finally, the generator model learns to fool the discriminator and to minimize the loss between the denoised image and the corresponding groundtruth.}
\end{figure*}

\begin{figure*}[ht]
    \includegraphics[width=0.9\textwidth]{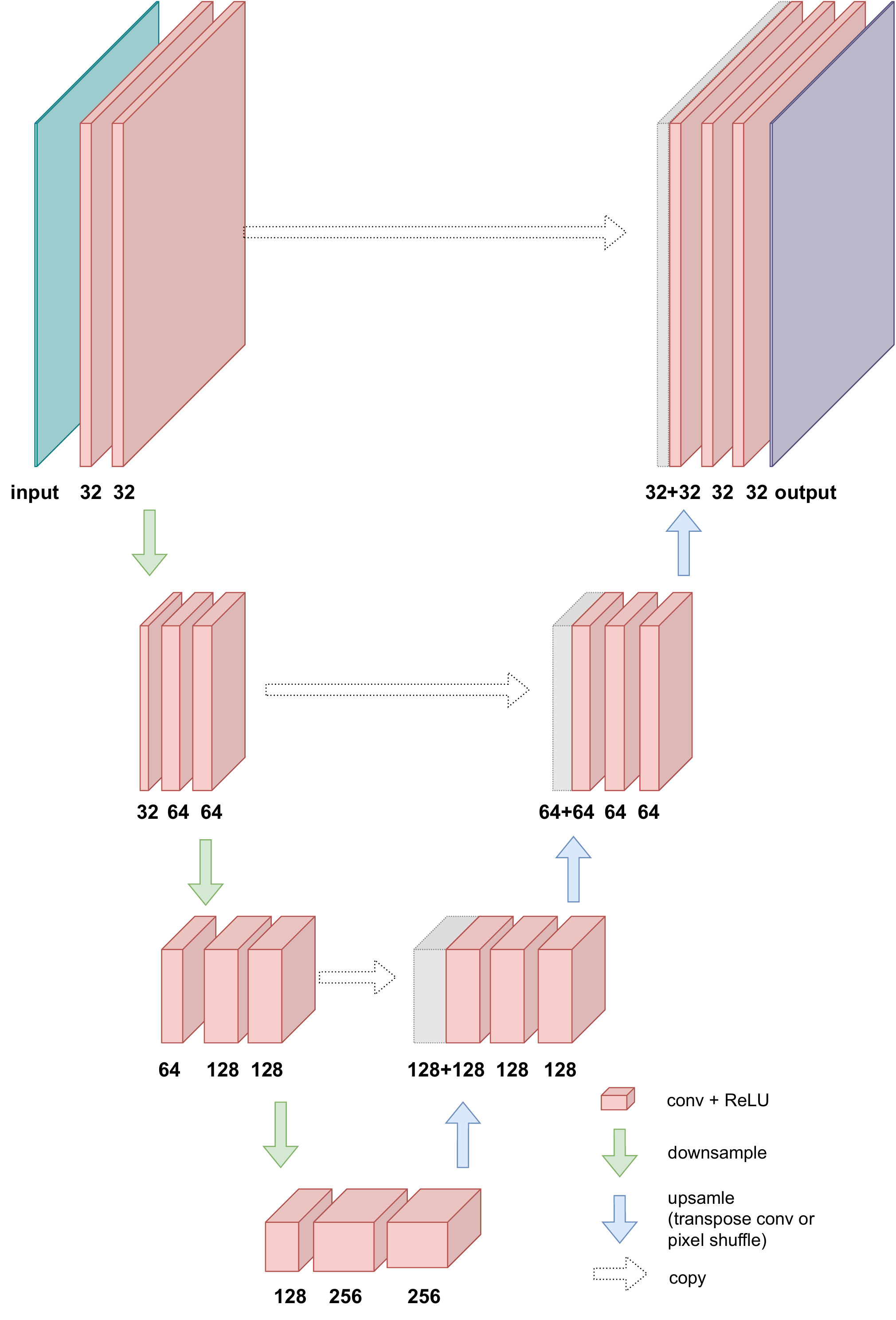}
     \caption{UNet architecture used for the experiments.  For upscaling we used pixel shuffle layer.}
\end{figure*}

\begin{figure*}[ht]
    \includegraphics[width=0.9\textwidth]{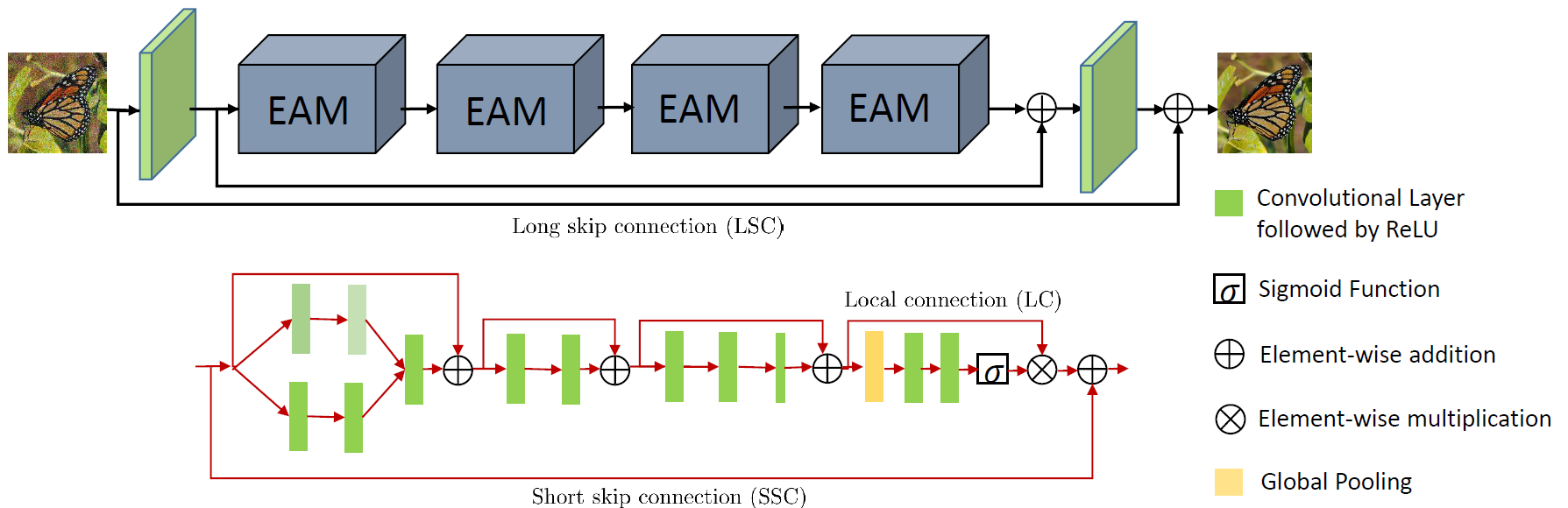}
      \caption{RIDNet architecture used for the experiments.}
\end{figure*}